# Electron recombination in low-energy nuclear recoils tracks in liquid argon

**Mariusz Wojcik**

*Institute of Applied Radiation Chemistry, Lodz University of Technology,
Wroblewskiego 15, 93-590 Lodz, Poland*
*E-mail*: mariusz.wojcik@p.lodz.pl

ABSTRACT: This paper presents an analysis of electron-ion recombination processes in ionization tracks of recoiled atoms in liquid argon (LAr) detectors. The analysis is based on the results of computer simulations which use realistic models of electron transport and reactions. The calculations reproduce the recent experimental results of the ionization yield from 6.7 keV nuclear recoils in LAr. The statistical distribution of the number of electrons that escape recombination is found to deviate from the binomial distribution, and estimates of recombination fluctuations for nuclear recoils tracks are obtained. A study of the recombination kinetics shows that a significant part of electrons undergo very fast static recombination, an effect that may be responsible for the weak drift-field dependence of the ionization yield from nuclear recoils in some noble liquids. The obtained results can be useful in the search for hypothetical dark matter particles and in other studies that involve detection of recoiled nuclei.

KEYWORDS: Ionization and excitation processes; Noble liquid detectors (scintillation, ionization, double-phase); Simulation methods and programs.

**Contents**

## 1. Introduction

Perhaps the only feasible way to detect the weakly interacting massive particles (WIMPs), expected to be a possible form of dark matter, is to look for an evidence of their elastic collisions with atomic nuclei on the Earth [1-3]. The nuclei recoiled in such collisions should have energy in the low keV range and produce dense ionization tracks of ~100 nm length in condensed media. Detection of nuclear recoils tracks is of interest not only for the dark matter search but also for studies on the postulated coherent elastic neutrino-nucleus scattering [4]. Particularly suitable target media for detecting low-energy ionization tracks are liquefied noble gases, mainly argon and xenon [5, 6]. They enable efficient extraction of electrons from the tracks by an applied electric field, and also have favorable scintillation properties. A number of experimental studies of low-energy nuclear recoils produced by monoenergetic neutron sources were done in the last decade using liquid xenon detectors [7-10]. Less results are available for liquid argon (LAr) [11, 12], but the interest in this target medium is increasing due to its advantages over other noble liquids, like the lower cost or good particle discrimination capabilities. Recently, the first results of the ionization yield of low-energy (6.7 keV) nuclear recoils in LAr were reported [13]. Very recently, the results obtained for nuclear recoils of energy from 16.9 to 57.3 keV have also been published [14].

The physical processes taking place in nuclear recoils tracks in condensed matter are still rather poorly understood. Although a general view of the mechanism of energy loss by recoiled atoms is provided by the theories originating from the Lindhard model [15-17], a more detailed information about track structure can only be obtained from simulations which suffer from serious uncertainties. Even less is known about the electron-ion recombination process. This is very disadvantageous, as the recombination probability directly affects both the ionization and scintillation yields. The available theories of Onsager [18], Jaffe [19], or Thomas and Imel [20] are for many reasons inapplicable to electron recombination in low-energy nuclear recoils tracks, especially in noble liquids where the electron transport cannot be described as ideal diffusion. In such a situation, empirical models are in use, with fitting parameters that usually have no physical significance. The lack of a realistic recombination model leaves many questions unexplained. For example, it is not clear why the ionization yield in LAr significantly depends on the applied electric field [13, 14], while almost no field effect is observed in liquid xenon detectors [7, 10].

Another important issue is recombination fluctuations. In the search for dark matter particles, one looks for the signatures of extremely rare events. The expected count rates for nuclear recoils originating from WIMPs collisions are on the order of $10^{-1}$ counts/kg/year for both argon and xenon [6]. To analyze so rare events, one needs to know not only the average recombination probability, but also the probability distribution of the number of recombined (or escaped) electrons.

This paper presents an analysis of electron-ion recombination in nuclear recoils tracks in LAr based on the results of computer simulations. This analysis gives an insight into the mechanism of the recombination process and provides quantitative predictions regarding the ionization yield and its fluctuations. The main element of the applied simulation method is the model of electron transport in LAr developed by Wojcik and Tachiya [21]. This model is based on the Cohen-Lekner theory of hot electrons in liquids [22], which postulates that the electron momentum and energy are exchanged with the liquid medium at different rates. In the model of Wojcik and Tachiya, these electron interactions are represented by two types of events – momentum-transfer collisions and energy-transfer collisions – both characterized by energy-dependent cross sections. Importantly, this model reproduces the electric field dependence of electron mobility in LAr [21]. The model of Wojcik and Tachiya was originally used to interpret the experimental data on bulk electron-ion recombination in LAr [23, 24], which could not be explained by the Debye-Smoluchowski theory. In recent years, our simulation model has also found successful applications related to detector physics. It was used to explain the LET dependence of the recombination factor measured in high-energy stopping muon and stopping proton tracks [25], and was used by the ArgoNeuT collaboration to interpret their results of the angular dependence, with respect to the drift field, of the recombination probability [26]. Our model was also used by Foxe *et al.* [27] to describe the recombination stage in their calculations of the ionization yield from electron recoils in LAr. Very recently, a study of nuclear recoils in LAr by those authors, independent of our work, has also appeared [28], which includes a limited description of the recombination process.

## 2. Method of calculation

To model electron recombination in a nuclear recoil track, we first need to know the detailed track structure, namely, the number and positions of ionized atoms. This information can in principle be obtained from track simulations. However, as already mentioned, such simulations suffer from large uncertainties originating, for example, from incomplete cross section data. Besides, they produce a variety of track structures from which it is hard to select the most representative ones. To circumvent these difficulties, we decided to use a different approach. We consider two limiting cases: (a) a linear track where all ionized atoms are placed on a straight line, and (b) a 'random' track where the position of each ionization is generated in a random direction from the previous ionization. The structures of real nuclear recoils tracks most probably fall in between the cases (a) and (b). In the simulations for both case (a) and case (b), the distance between successive ionizations is randomly picked from an exponential distribution with the mean value $r_{ion}$. According to the review paper by Chepel and Araujo [6] (see the data compilation in their table 1), the LET of low-energy (5 keV) nuclear recoils in LAr is about $1.9\times10^3$ MeV/(g·cm$^{-2}$), with the electronic part being estimated as ~0.2 of this value. Hence the electronic part of LET is approximately equal to 50 eV/nm. Using the widely accepted value of the effective energy spent on producing one ionization in LAr, $W_{eff} = 23.6$ eV [29], we obtain the mean distance between successive ionizations of about 0.5 nm. Therefore, in the calculations described in this paper we assumed $r_{ion} = 0.5$ nm. Concerning the initial number of ions in the modeled tracks, we used three values, $N_i = 30$, 50, and 70, which correspond to the nuclear recoil energy between about 3.5 and 8 keV and thus cover an energy range important for the dark matter searches.

The simulation method applied in the present study to model the electron recombination processes is similar to that used in refs. [25, 26]. The electrons are randomly generated at a distance $r_0$ from their parent cations and assigned an initial kinetic energy $E_{k0}$. Then, the classical electron trajectories in the total electric field, which includes the Coulomb fields of all other electrons and cations and an applied external field $F$, are calculated. At each time step $\Delta t$, the electron collision probabilities are determined as

$$P_j = 1 - \exp(-\Delta t/\tau_j), \quad j = 1, 2, \ldots, N, \tag{1}$$

where the mean free times $\tau_j$ are related to the collision cross sections [21] and $N$ is the number of electrons. The momentum-transfer collisions are modeled as isotropic elastic electron-atom collisions, while the energy-transfer collisions change only the magnitude of electron velocity without changing its direction. The electron trajectories and collisions are simulated until all electrons disappear due to either recombination or escape. By recording these events and repeating the simulation for many independently generated tracks, the recombination (or escape) probability can be obtained. The present application of our simulation model differs from its previous applications [25, 26] in two aspects. First, all cations and electrons forming a nuclear recoil track are now explicitly modeled, without any track periodicity assumed. Second, the cations are no longer immobile. They are initially assigned velocities sampled from the Maxwell-Boltzmann distribution, and their trajectories in the total electric field are then calculated. The cations undergo collisions with the mean free time $\tau_c = \mu_c m_c / e$, where $\mu_c$ and $m_c$ are the cation mobility and mass, respectively, and $e$ is the elementary charge. The collision probability at each time step is determined using an expression analogous to eq. (1). In each collision, the cation velocity is randomized to restore the Maxwell-Boltzmann distribution (the collisions are isotropic). This procedure properly models both the cation drift and diffusion. Other details of the simulation method are widely described in ref. [25].

The calculations were carried out for temperature $T = 87$ K, density $n = 2.11 \times 10^{22}$ cm$^{-3}$, and the dielectric constant $\varepsilon = 1.49$. The cation mobility $\mu_c = 1.52 \times 10^{-3}$ cm$^2$V$^{-1}$s$^{-1}$ [30] was assumed. The space-energy criterion of recombination was used, as described in ref. [25], with the critical distance $r_{crit} = 1.29$ nm and energy $E_{k,crit} = 1$ eV (this form of the recombination criterion allowed us earlier to reproduce the measured electron escape probability from high-energy particle tracks in LAr [25]). The electrons were assumed to escape recombination when they got separated from the center of the initial track structure by more than $r_{max} = 2500$ nm. In the preliminary stage of calculations, extensive tests were performed to determine sensitivity of the simulation results to parameter changes. The calculated escape probability was found to very weakly (<1%) depend on $r_{crit}$, $E_{k,crit}$, and $r_{max}$ over the test ranges of 1-1.5 nm, 0.5-1.5 eV, and 2500-5000 nm, respectively. The effect of cation mobility was also found to be small (the escape probability increased by 1-2% when $\mu_c$ was set to zero). Unexpectedly, the dependence of the simulation results on the initial electron energy $E_{k0}$ and, in consequence, the initial distance $r_0$, turned out to be much stronger. For example, the change of $E_{k0}$ from 5 to 8 eV (with $r_0 = 0.5$ nm) was found to increase the escape probability by about 50%. This behavior is different from that observed in the earlier study of electron thermalization in LAr [31], carried out for a single electron-ion pair, where practically no effect of the initial electron energy was found over the range of several eV. Since no data are available that describe the energy of electrons ejected from ionized atoms in nuclear recoils tracks, we decided to initialize the electron energy in the simulation using a theoretical energy distribution of secondary electrons for low-energy electron collisions with argon atoms obtained from the EEDL database [32]. The following procedure was applied. For each initialized electron, an energy $E_0$ was sampled from the distribution that describes particles ejected from the M3 shell of Ar by 3.98 keV primary electrons. Then, the initial kinetic energy was calculated as $E_{k0} = E_0 + e^2/4\pi\varepsilon_0\varepsilon r_0$, with $r_0 = 0.2$ nm being assumed as the cation radius. An attempt was also made to assess how the simulation results would be affected by the change of the electron energy distribution resulting from interactions in the liquid phase. This was done by taking into account the conduction band energy $V_0$ in the initialized kinetic energy of electrons, as will be described later.

## 3. Results

Figure 1 shows the average number of escaped electrons $N_e$ as a function of applied field. These results were obtained from the simulations using both the linear (closed circles) and random (open circles) models of the track with $N_i = 30$, 50, and 70. The results depend on the track model used, with 20-30% less electrons escaping from the random tracks than from the corresponding linear tracks. The shape of the field dependence of $N_e$ is very similar for both types of track. However, the field effect is becoming stronger as the initial number of ions increases. The results shown in figure 1 by triangles were obtained for the linear tracks of 70 ions using a modified method of generating the initial electron energy, in which $E_0$ was increased by $-V_0 = 1$ eV. Various estimates of the electron energy at the bottom of the conduction band in LAr exist [33], usually in the range from 0 to -0.3 eV (with respect to the vacuum level). By setting $V_0 = -1$ eV we tried to assess the maximum effect that the use of a liquid-phase electron energy distribution, if available, would have on the simulation results. As seen from figure 1, this effect should not be strong, as only about 7% more electrons escape from the track when $V_0$ is set to -1 eV (very similar results, not shown in figure 1, were obtained for other values of $N_i$, and for the random tracks).

In addition to the simulation results, figure 1 shows the recent experimental data of Joshi *et al.* [13] for the number of detected electrons from 6.7 keV nuclear recoils in LAr. These authors estimate the total number of ionizations in the observed tracks as 72±2. Although our estimation of the track parameters, presented in section 2, suggests that this number might be somewhat lower (~60), it seems to be justified to compare the data of Joshi *et al.* with the simulation results obtained for $N_i = 70$. Figure 1 shows that the experimental data are quite well reproduced by the simulation when the linear model of track is used. Although the calculated field dependence of $N_e$ is weaker than that observed in experiment, the simulation results are mostly within the experimental uncertainties. The results of the random track model are well below the experimental results (except for the lowest drift field). This is probably due to the fact that the random model represents a theoretical limit rather than a typical structure of the tracks. It is hard to imagine that a significant part of nuclear recoils tracks do not have any orientational correlation with the direction of motion of the primary recoil. The linear track model seems to better represent the shape of real tracks, and an application of this model in simulation gives reasonable agreement with experiment.

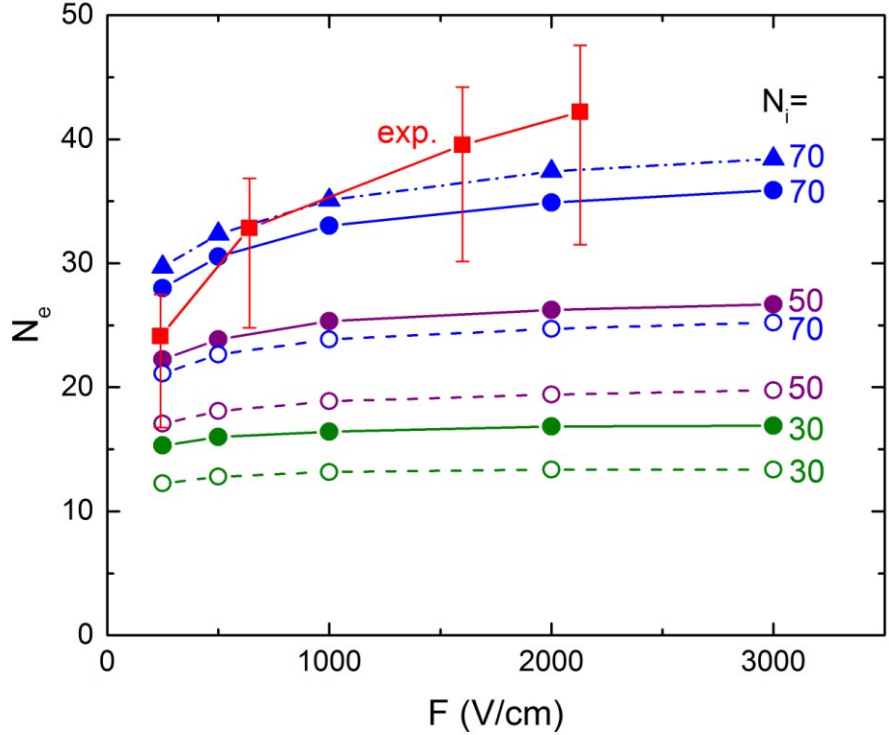

**Figure 1.** Mean number of escaped electrons as a function of applied drift field. The simulation results obtained for $N_i = 30$, 50, and 70 using the linear (closed circles) and random (open circles) models of the track. The results shown by triangles were calculated for the linear tracks using the modified electron energy distribution with $V_0 = -1$ eV, as described in the text. The squares show the experimental results of Joshi *et al.* [13] obtained for 6.7 keV nuclear recoils in LAr ($N_i = 72\pm2$).

In figure 2 the simulation results obtained using the linear track model are shown as the electron escape probability $P_{esc} = N_e/N_i$ (often termed the recombination factor). $P_{esc}$ is seen to decrease with increasing $N_i$, which can be explained as a result of stronger Coulomb attraction in tracks formed of a larger number of cations and electrons. We tried to describe these results using the Thomas-Imel box model [20]

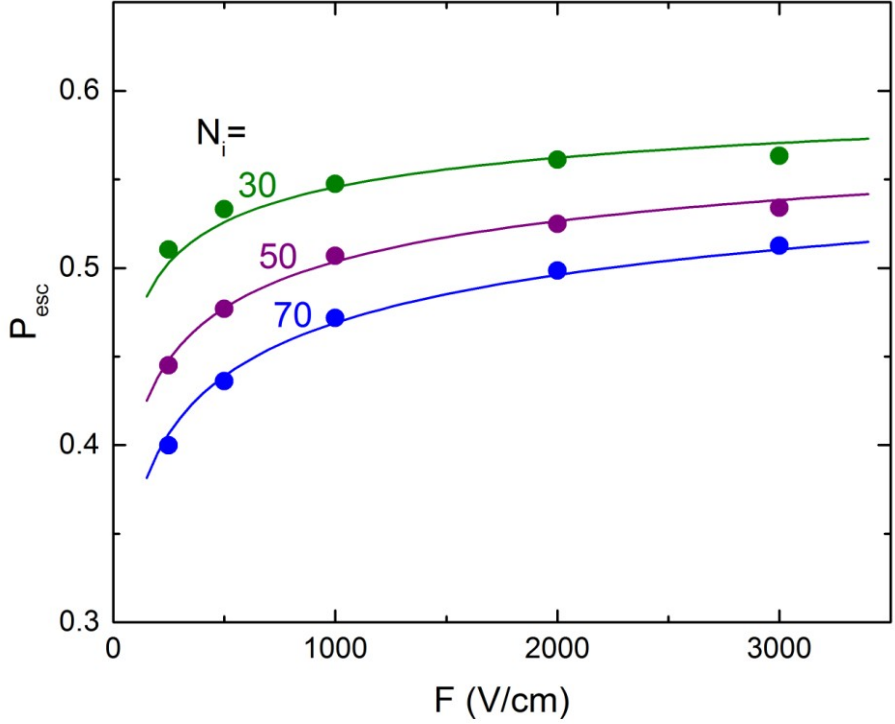

**Figure 2.** Electron escape probability as a function of drift field. The circles represent the simulation results obtained for $N_i = 30$, 50, and 70 using the linear track model. The curves show the global fit of equation (3) to the simulation data.

$$P_{esc} = \frac{1}{\xi}\ln(1+\xi), \qquad \xi = \frac{N_i C}{F}, \qquad (2)$$

where $C$ is constant, but the field dependence obtained from the simulation is much weaker than that expressed by equation (2). Good fits were obtained when an empirical modification of equation (2) was applied with $F$ being replaced by $F^b$ [34]. However, the fitted parameters were then found to strongly depend on $N_i$, with $b$ and $C$ changing from 0.10 to 0.22 and from 0.13 to 0.19, respectively, as $N_i$ was increased from 30 to 70.

Looking for a better empirical model, we studied the kinetics of the recombination process. Figure 3 shows the time dependence of the mean number of survived electrons $N(t)$ calculated for the linear tracks of 30, 50, and 70 ions, and for the electric fields of 250 V/cm (solid lines) and 3000 V/cm (dashed lines). An interesting effect revealed by these data is that a significant part of simulated electrons undergo recombination at very short times. This effect is independent of the applied field. To exclude the possibility of a simulation artifact that might be related, for example, to the use of a space-energy recombination criterion, we ran test calculations with an additional restriction that recombination is not allowed at times shorter than a certain critical time $t_{crit}$. The use of this restriction with $t_{crit} = 10^{-12}$ s was found to increase $P_{esc}$ at $N_i = 70$ by only 1.4% at 250 V/cm and 4.4% at 3000 V/cm. The effects of varying the other critical values, $r_{crit}$ and $E_{k,crit}$, were also found to be weak, as already discussed. It can therefore be postulated that the very fast recombination of a significant part of electrons is a real effect, most probably caused by electron trapping in the Coulomb fields from densely located cations. This effect can be called the initial or static recombination. The electrons involved in this process do not separate from the cations to larger distances nor undergo the usual thermalization. They remain in close orbits or Rydberg-like states around the cations and quickly recombine. The probability of static recombination for a particular electron should be the lower, the higher its initial energy $E_0$. This was confirmed by a detailed analysis of the simulation results. However, this probability was found to be significant (~15%) even for those electrons which were assigned high initial energies $E_0 > 5$ eV.

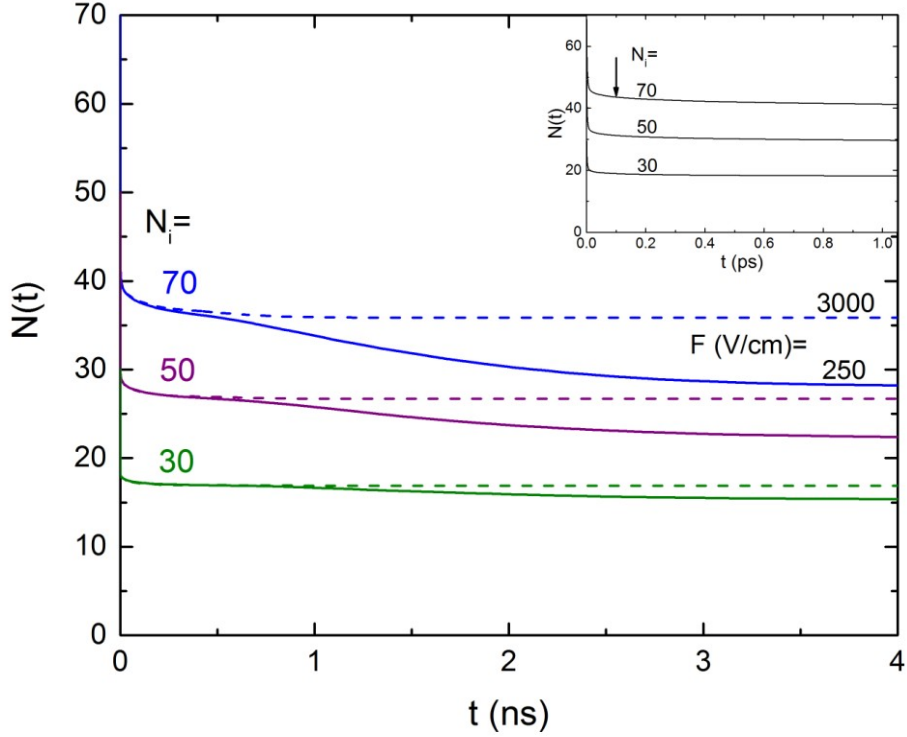

**Figure 3.** Mean number of survived electrons as a function of time. The simulation results obtained for $F = 250$ V/cm (solid curves) and 3000 V/cm (dashed curves). The inset shows the short-time recombination kinetics. It is assumed that the static recombination stage ends at $t = 10^{-13}$ s, as indicated by the arrow.

The fraction of electrons that avoid static recombination, determined at $t = 10^{-13}$ s (see the inset to figure 3), was found to be practically independent of $N_i$ and equal to 0.63±0.01. Denoting this fraction by $\alpha$, we propose a further modification of the Thomas-Imel box model in the following form

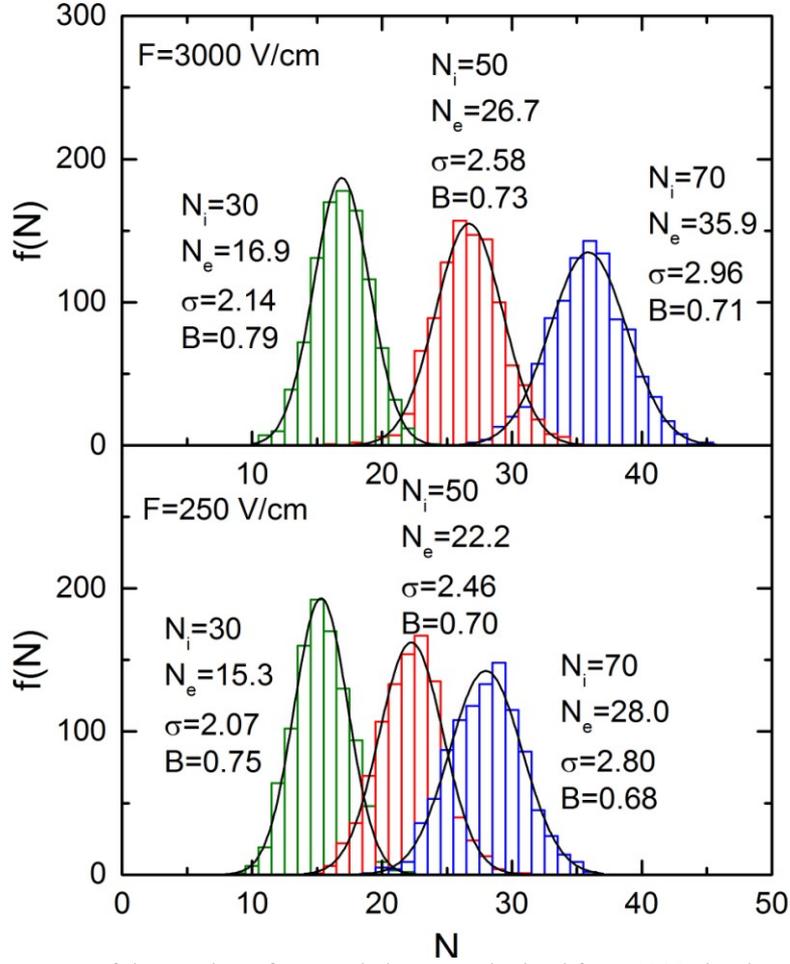

**Figure 4.** Histograms of the number of escaped electrons obtained from 1000 simulated tracks at $F = 250$ V/cm (lower panel) and 3000 V/cm (upper panel). The solid curves show approximations of the histograms by normal distributions. The parameters $N_i$, $N_e$, $\sigma$, and $B$ corresponding to each histogram are shown in the figure.

$$P_{esc} = \frac{\alpha}{\xi}\ln(1+\xi), \qquad \xi = \frac{\alpha N_i C}{F^b}. \tag{3}$$

Equation (3) is based on an assumption that the earlier empirical model, proposed in Ref. [34], properly describes the recombination of those electrons that survive the static recombination. With $\alpha$ fixed to 0.63, we performed a two-parameter global fitting of equation (3) to all simulation data plotted in figure 2, and obtained $C = 0.227$ and $b = 0.374$ (when $F$ is expressed in V/cm) with the standard errors of 0.022 and 0.014, respectively. As shown in figure 2, the fitted curves are in good agreement with the simulation results, considering their dependence on both $F$ and $N_i$.

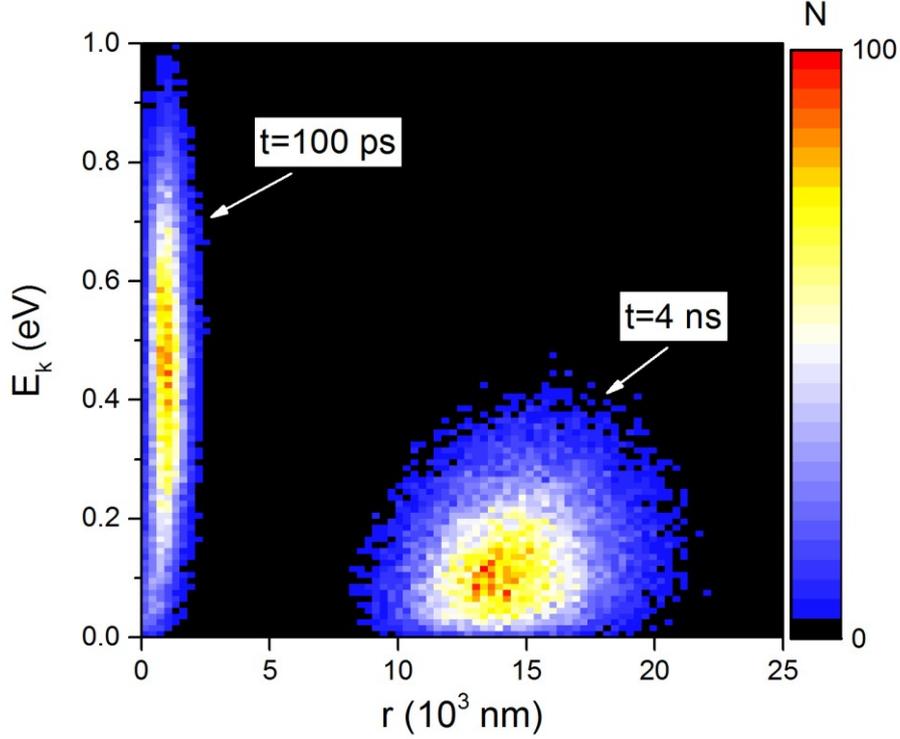

**Figure 5.** Histograms of electron kinetic energy and distance from the track center at $t = 100$ ps and 4 ns. The simulation results obtained for $N_i = 70$ and $F = 3000$ V/cm. The height of the histogram at $t = 100$ ps is reduced by the factor of two.

We now turn to an analysis of the statistical distribution $f(N)$ of the number of electrons that escape recombination. Figure 4 shows the histograms of $N$ calculated for $N_i = 30$, 50, and 70 using the linear track model. Each histogram was obtained from 1000 independently simulated tracks. If the fate of each electron in a track (recombination vs. escape) was independent of the fates of other electrons in the same track, the distribution $f(N)$ should be binomial. At $N_i \geq 30$ this binomial distribution should in turn be well approximated by the normal distribution with the mean

$$\mu_B = N_e = P_{esc} N_i \tag{4}$$

and the standard deviation

$$\sigma_B = \sqrt{P_{esc}(1 - P_{esc}) N_i} \; . \tag{5}$$

As seen from figure 4, the shapes of the histograms are well described by normal distributions. The standard deviations obtained from these histograms, $\sigma = \sqrt{\langle (N - N_e)^2 \rangle}$, increase with increasing $N_i$. However, the values of $\sigma$ are found to be smaller than the corresponding values of $\sigma_B$. We quantify this effect using a parameter $B$ defined by

$$B = \frac{\sigma}{\sigma_B} . \qquad (6)$$

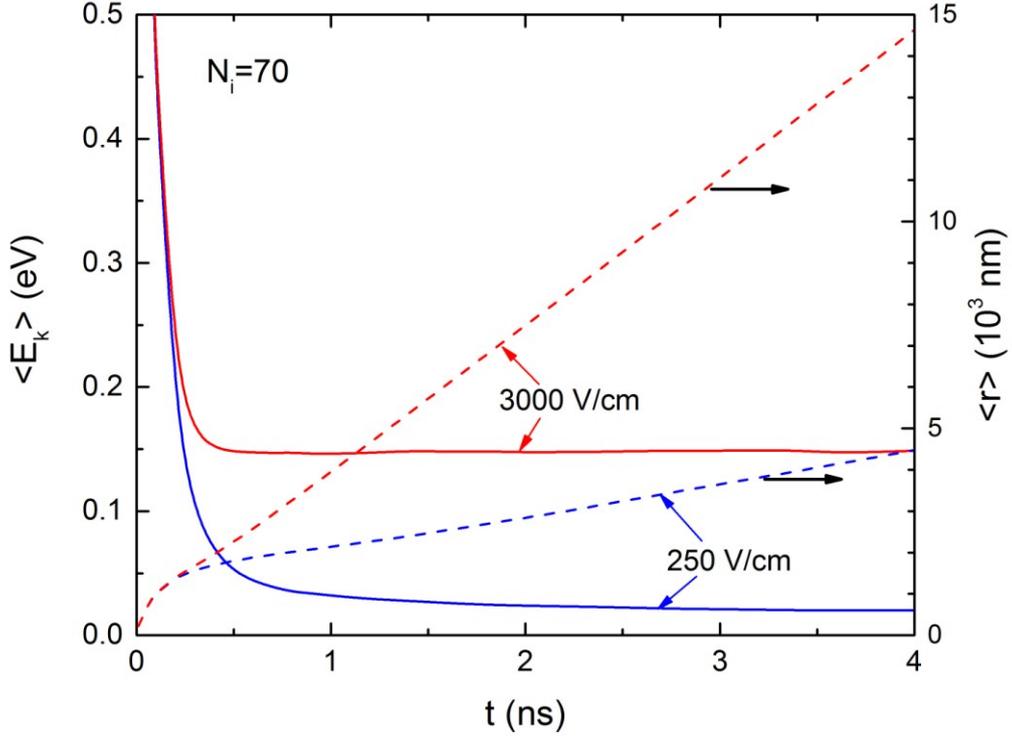

**Figure 6.** The average electron kinetic energy (solid curves, left scale) and the average electron distance from the track center (dashed curves, right scale) as functions of time. The simulation results obtained for $N_i = 70$ and $F = 250$ and 3000 V/cm.

For all histograms shown in figure 4, $B$ is smaller than unity. This deviation from the binomial distribution is becoming larger as the number of ions increases. When $N_i$ is changed from 30 to 70, $B$ decreases from 0.75 to 0.68 at $F = 250$ V/cm, and from 0.79 to 0.71 at $F = 3000$ V/cm. This behavior can be understood as a result of stronger mutual interactions in tracks formed of a larger number of particles. The interparticle correlations become relatively weaker as the drift field increases, which explains why the values of $B$ are larger at 3000 V/cm than at 250 V/cm.

The simulation methodology applied in this study can also give us an insight into the electron thermalization process. An example is given in figure 5, which shows two-dimensional distributions of electron kinetic energy $E_k$ and distance $r$ from the track center obtained at $t = 100$ ps and 4 ns. The average energy $\langle E_k \rangle$ and distance $\langle r \rangle$ as functions of time are plotted in figure 6 ($N_i = 70$, $F = 250$ and 3000 V/cm). $\langle E_k \rangle$ is seen to decrease to an equilibrium energy, which is strongly dependent on the drift field and significantly higher than the thermal energy $3k_B T/2 = 0.011$ eV (so, strictly speaking, this process should not be called thermalization). At $F = 250$ V/cm the electron energy equilibrates at $t \sim 2$ ns. This time is approximately equal to the mean thermalization time of 1.8 ns obtained in ref. [31]. However, at $F = 3000$ V/cm the equilibration occurs much earlier (~0.5 ns). Also the mean distance $\langle r \rangle$ significantly depends on the applied field, except for times shorter than ~0.3 ns. Therefore, the mean thermalization distance of 2600 nm [31] does not correctly describe the electron thermalization in nuclear recoils tracks at strong drift fields.

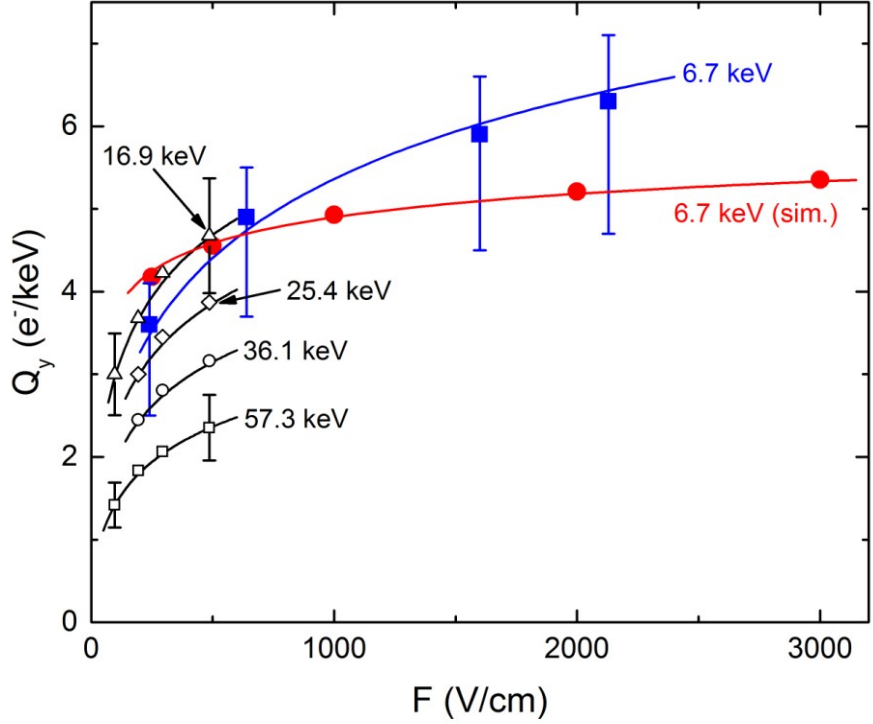

**Figure 7.** The ionization yield in [e⁻/keV] as a function of applied drift field. The open symbols show the experimental data of Cao *et al.* [14] obtained for four nuclear recoil energies from 16.9 to 57.3 keV, as indicated in the figure (for the sake of clarity, the error bars are shown only for selected data points). The results of Joshi *et al.* [13] are represented by closed squares. The closed circles show the simulation results obtained for $N_i = 70$ using the linear track model (cf. figure 1) with the ionization yield being calculated as $Q_y = N_e/E$ where $E = 6.7$ keV.

## 4. Discussion

In this paper we have given an analysis of various aspects of electron recombination in low-energy nuclear recoils tracks in LAr. Our analysis is based on the simulation model which was previously applied to different ionization phenomena in LAr yielding reasonable agreement with experiment. Also in the present study, the obtained results reproduce the experimental data quite well (see figure 1). It should be noted that this agreement has been achieved without parameter fitting. On the other hand, the applied simulation model, as any other theoretical model, involves approximations and is not free from systematic errors. It is very difficult to assess the overall magnitude of these errors, although extensive tests have been performed to determine sensitivity of the simulation results to parameter changes, as described earlier. The statistical simulation errors in all of the presented results are very small (<1%).

It is instructive to compare the results of the present work with the experimental data obtained not only for nuclear recoils of energy $E = 6.7$ keV, but also for those of higher energies. In figure 7 we plot the ionization yield $Q_y$, defined as the number of escaped electrons per unit (1 keV) nuclear recoil energy, as a function of applied field. The open symbols show the results of Cao *et al.* [14] obtained for four nuclear recoil energies between 16.9 and 57.3 keV. $Q_y$ determined by these authors decreases with increasing $E$ at all applied fields. This trend is well reproduced by the simulation results presented in figure 2, noting that the escape probability $P_{esc}$ is proportional to $Q_y$, and the number of ions $N_i$ is proportional to $E$ (it is assumed here that the average energy spent on producing one ionization in a track is approximately independent of the nuclear recoil energy).

The ionization yield measured by Joshi *et al.* [13] at $E = 6.7$ keV (shown by closed squares in figure 7) is lower than that obtained by Cao *et al.* at $E = 16.9$ keV. This unexpected behavior most probably results from experimental uncertainties. As shown by the error bars included in figure 7, these uncertainties can be as large as 15-30%.

We now discuss the drift-field dependence of the ionization yield. As shown in figure 2, the simulation predicts this dependence to become stronger as the number of ions increases. This trend is also confirmed by the experimental results of Cao *et al.*. The ratio $Q_y(F = 486 \text{ V/cm})/Q_y(F = 96.4 \text{ V/cm})$ calculated using their data is 1.56 at $E = 16.9$ keV, and increases to 1.66 at $E = 57.3$ keV.

While the simulation results reproduce the trends observed experimentally, the calculated drift-field dependence of $Q_y$ at $E = 6.7$ keV is weaker than that measured by Joshi *et al.* [13]. Due to the large uncertainties of the experimental results, we are not completely sure which data better describe the electron recombination in LAr. It was postulated in this paper that a weak field dependence of the escape probability might result from the fast, static recombination which cannot be prevented even by a strong external field. Although this effect is a theoretical prediction that needs to be further verified, it is supported by the results of simulation tests described earlier. Even if not confirmed in LAr, the static recombination might explain the very weak drift-field dependence of the charge yield obtained from the measurements of ionization from nuclear recoils in liquid xenon [7, 10]. As demonstrated by Manzur *et al.* [10] (see their table I), almost no suppression of electron-ion recombination by the applied field was observed over a wide range of nuclear recoil energy between 3.9 and 66.7 keV. While the simulation model used in the present study is not directly applicable to xenon, the mechanisms of electron recombination in these two noble liquids might not be very different.

This paper provides an estimation of recombination fluctuations in low-energy nuclear recoils tracks in LAr. The calculated fluctuations increase with increasing nuclear recoil energy, but are smaller than those obtained from the binomial distribution. No experimental data on recombination fluctuations in LAr are currently available that could be compared with our simulation results. In the case of liquid xenon, analyses of experimental data [35, 36] led to the conclusion that the recombination fluctuations are larger than those given by the binomial distribution. It should be noted, however, that this conclusion was based on the results obtained mostly for electron recoils. The situation might be different in nuclear recoils tracks, where the ionization density is substantially higher and the interparticle correlations are therefore much stronger.

We hope the proposed simulation methodology and the analysis of electron recombination presented in this paper will be useful in the search for dark matter particles or in other studies that involve observation of recoiled nuclei. A generalization of our simulation model to liquid xenon is worth to be undertaken.

## Acknowledgment

This work was supported by the National Science Centre of Poland (Grant No. DEC-2013/09/B/ST4/02956).

## References

[1] G. Bertone, D. Hooper and J. Silk, *Particle dark matter: evidence, candidates and constraints*, *Phys. Rep.* **405** (2005) 279.

[2] R.J. Gaitskell, *Direct detection of dark matter*, *Annu. Rev. Nucl. Part. Sci.* **54** (2004) 315.

[3] S. Ahlen *et al.*, *The case for a directional dark matter detector and the status of current experimental efforts*, *Int. J. Mod. Phys. A* **25** (2010) 1.

[4] A. Drukier and L. Stodolsky, *Principles and applications of a neutral-current detector for neutrino physics and astronomy*, *Phys. Rev. D* **30** (1984) 2295.

[5] E. Aprile, A.E. Bolotnikov, A.I. Bolozdynya and T. Doke, *Noble Gas Detectors,* Wiley-VCH, Weinheim 2006.

[6] V. Chepel and H. Araujo, *Liquid noble gas detectors for low energy particle physics*, 2013 *JINST* **8** R04001.

[7] E. Aprile *et al.*, *Simultaneous measurement of ionization and scintillation from nuclear recoils in liquid xenon for a dark matter experiment*, *Phys. Rev. Lett.* **97** (2006) 081302.

[8] V. Chepel *et al.*, *Scintillation efficiency of liquid xenon for nuclear recoils with the energy down to 5 keV*, *Astropart. Phys.* **26** (2006) 58.

[9] E. Aprile *et al.*, *New measurement of the relative scintillation efficiency of xenon nuclear recoils below 10 keV*, *Phys. Rev. C* **79** (2009) 045807.

[10] A. Manzur *et al.*, *Scintillation efficiency and ionization yield of liquid xenon for monoenergetic nuclear recoils down to 4 keV*, *Phys. Rev. C* **81** (2010) 025808.

[11] D. Gastler *et al.*, *Measurement of scintillation efficiency for nuclear recoils in liquid argon*, *Phys. Rev. C* **85** (2012) 065811.

[12] T. Alexander *et al.*, *Observation of the dependence on drift field of scintillation from nuclear recoils in liquid argon*, *Phys. Rev. D* **88** (2013) 092006.

[13] T.H. Joshi *et al.*, *First measurement of the ionization yield of nuclear recoils in liquid argon*, *Phys. Rev. Lett.* **112** (2014) 171303.

[14] H. Cao *et al.*, *Measurement of scintillation and ionization yield and scintillation pulse shape from nuclear recoils in liquid argon*, *Phys. Rev. D* **91** (2015) 092007.

[15] J. Lindhard, M. Scharff and H.E. Schiøtt, *Range concepts and heavy ion ranges*, *Mat. Fys. Medd. Dan. Vid. Selsk.* **33** (1963) 1.

[16] A. Mangiarotti *et al.*, *A survey of energy loss calculations for heavy ions between 1 and 100 keV*, *Nucl. Instrum. Meth. A* **580** (2007) 114.

[17] P. Sorensen, *Atomic limits in the search for galactic dark matter*, *Phys. Rev. D* **91** (2015) 083509.

[18] L. Onsager, *Initial recombination of ions*, *Phys. Rev.* **54** (1938) 554.

[19] G. Jaffe, *Zur Theorie der Ionisation in Kolonnen*, *Ann. Phys.* **42** (1913) 303.

[20] J. Thomas and D.A. Imel, *Recombination of electron-ion pairs in liquid argon and liquid xenon*, *Phys. Rev. A* **36** (1987) 614.

[21] M. Wojcik and M. Tachiya, *Electron transport and electron–ion recombination in liquid argon: simulation based on the Cohen–Lekner theory*, *Chem. Phys. Lett.* **363** (2002) 381.

[22] M.H. Cohen and J. Lekner, *Theory of hot electrons in gases, liquids, and solids*, *Phys. Rev.* **158** (1967) 305.

[23] K. Shinsaka, M. Codama, T. Srithanratana, M. Yamamoto and Y. Hatano, *Electron-ion recombination rate constants in gaseous, liquid, and solid argon*, *J. Chem. Phys.* **88** (1988) 7529.

[24] M. Wojcik and M. Tachiya, *Electron-ion recombination in dense gaseous and liquid argon: effects, due to argon cation clusters allow to explain the experimental data*, *Chem. Phys. Lett.* **390** (2004) 475.

[25] M. Jaskolski and M. Wojcik, *Electron recombination in ionized liquid argon: A computational approach based on realistic models of electron transport and reactions*, *J. Phys. Chem. A* **115** (2011) 4317.

[26] R. Acciarri *et al.*, *A study of electron recombination using highly ionizing particles in the ArgoNeuT Liquid Argon TPC*, 2013 *JINST* **8** P08005.

[27] M. Foxe *et al.*, *Low-energy (<10 keV) electron ionization and recombination model for a liquid argon detector*, *Nucl. Instrum. Meth. A* **771** (2015) 88.

[28] M. Foxe *et al.*, *Modeling ionization and recombination from low energy nuclear recoils in liquid argon*, *Astropart. Phys.* **69** (2015) 24.

[29] M. Miyajima *et al.*, *Average energy expended per ion pair in liquid argon*, *Phys. Rev. A* **9** (1974) 1438.

[30] N. Gee, S.S.-S. Huang, T. Wada and G.R. Freeman, *Comparison of transition from low to high density transport behavior for ions and neutral molecules in simple fluids*, *J. Chem. Phys.* **77** (1982) 1411.

[31] M. Wojcik and M. Tachiya, *Electron thermalization and electron–ion recombination in liquid argon*, *Chem. Phys. Lett.* **379** (2003) 20.

[32] D.E. Cullen, S.T. Perkins and S.M. Seltzer, *Technical Report UCRL-50400, Vol. 31,* Lawrence Livermore National Laboratory, November 1991; available from: https://www-nds.iaea.org/epdl97/.

[33] I.T. Steinberger, *Band structure parameters of classical rare gas liquids*, in: W. F. Schmidt and E. Illenberger (Eds.), *Electronic excitations in liquefied rare gases,* American Scientific Publishers 2005, Chapter 3.

[34] C.E. Dahl, *The physics of background discrimination in liquid xenon, and first results from XENON10 in the hunt for WIMP dark matter*. PhD thesis, Princeton University 2009.

[35] A. Dobi, *Measurement of the electron recoil band of the LUX dark matter detector with a tritium calibration source*. PhD thesis, University of Maryland 2014.

[36] B. Lenardo *et al.*, *A global analysis of light and charge yields in liquid xenon*, arXiv:1412.4417.